\documentclass[aps,prd,preprint,showpacs,nofootinbib]{revtex4}
\usepackage{epsfig}
\def\bea{\begin{eqnarray}}
\def\eea{\end{eqnarray}}
\begin{document}
\title{Soft Gluon Resummation Effects in Single Graviton Production
 at the CERN Large Hadron Collider in the Randall-Sundrum Model}
\author{\center{Qiang Li\footnote{\hspace{-0.1cm}
Electronics address: qliphy@pku.edu.cn}, Chong Sheng
Li\footnote{\hspace{-0.1cm} Electronics address: csli@pku.edu.cn},
and Li Lin Yang\footnote{\hspace{-0.1cm} Electronics address:
llyang@pku.edu.cn}}} \affiliation{\small Department of Physics,
Peking University, Beijing 100871, China}

\begin{abstract}
We study QCD effects in single graviton production at the CERN
Large Hadron Collider (LHC) in the Randall-Sundrum (RS) Model. We
present in detail the complete next-to-leading order (NLO) QCD
corrections to the inclusive total cross sections. The NLO QCD
corrections enhance significantly the total cross sections and
decrease efficiently the dependence of the total cross sections on
the factorization and renormalization scales. We also examine the
uncertainty of the total cross sections due to the parton
distribution function (PDF) uncertainties. For the differential
cross sections on the transverse momentum ($q_T$) of the graviton,
within the CSS resummation formalism, we resum the
logarithmically-enhanced terms at small $q_T$ to all orders up to
NLO logatithmic accuracy. Combined with the fixed order
calculations, we give consistent predictions for both small $q_T$
and large $q_T$.
\end{abstract}


\maketitle
\section{Introduction}
Now search for extra dimensions has been one of the major objects at
the LHC, since its physical effects can appear at the TeV energy
scale. The idea of extra dimensions was revived in the
1990's\cite{ADD,RS,lyk,witt,hora,anto}, which can bring new
solutions to the gauge hierarchy problem and be used to resolve some
problems of the SM such as the origin of the fermion masses and
their hierarchy.

So far, there have been various extra dimension models, which can
be divided into two major classes according to the geometry of the
background space-time manifold. The first one includes the ADD
model\cite{ADD} and its variants, which extend the dimension of
the totoal space-time to $D=4+\delta$, propose a factorizable
metric, and get large size of the extra dimensions ($\gg 1/M_{\rm
p}$). In the ADD model, the SM particles live in the usual
$4-$dimensional space-time, while gravity can propagate in the
additional $\delta$-dimensional space, which is assumed for
simplicity to be compactified on the $\delta$-dimensional torus
$T^\delta$ with a common radius R. Then the 4-dimensional Planck
scale $M_{\rm{p}}$ is related to the fundamental scale $M_s$ as
follows\cite{ADD,csaki}:
\begin{eqnarray}\label{scale}
M^2_{{\rm p}}=M^{\delta+2}_s(2\pi R)^\delta,
\end{eqnarray}
where $M_s\sim {\rm TeV}$. According to Eq.~(\ref{scale}),
deviations from the usual Newtonian gravitational force law can be
expected at distances smaller than $R\sim
2\times10^{-17}\times10^{\frac{32}{\delta}}{\rm cm}$\cite{csaki}.
For $\delta\geq 2$, ADD is consistent with the current
experiments\cite{grav} since gravitational forces are not yet well
probed at distances less than about a millimeter. However for
$\delta=2$, there are constraints arising from, e.g., supernova
cooling, which require $M_s\geq 10-100 {\rm TeV}$ if $\delta=2$
\cite{csaki}.

The second one includes the 5-dimensional RS model\cite{RS} and its
variants, in which a warped metric is introduced and the size of the
extra dimension needs not to be too large compared with the Planck
length. In the RS model, the extra dimension is assumed to be an
$S_1/Z_2$ orbifold, which has two fixed points, $\theta=0$ and
$\theta=\pi$. At each fixed point, there is a 3-brane, and the brane
at $\theta=\pi$ corresponds to the brane we live on, while the one
at $\theta=0$ is the high energy brane. Between the two 3-branes is
a slice of AdS space, where only the graviton can propagate into.
Moreover, the 4-dimensional metric is the function of the coordinate
of the 5th dimension, i.e.
\begin{equation}\label{rsm}
ds^2=e^{-2kr_c|\phi|}(\eta_{\mu\nu}+\frac{2}{M_*^{3/2}}h_{\mu\nu})dx^\mu
dx^\nu-r^2_cd\phi^2,\,\, 0\leq|\phi|\leq \pi, \end{equation} where
$k$ is a scale of order of the Planck scale and relates the
5-dimensional Planck scale $M_*$ to the cosmological constant, $r_c$
is the compactification radius, and $h_{\mu\nu}$ is the graviton.

After solving the 5-dimensional Einstein equation, we can get the
tensions of the two branes\cite{csaki}
\begin{equation}
V_0=-V_\pi=12 k M_*^3,
\end{equation}
and from Eq.(\ref{rsm}), we can get the relation between $M_*$ and
4-dimensional reduced Planck scale $\overline{M}_{P}$\cite{RS}
\begin{eqnarray}
\overline{M}_{P}^2=\frac{M_*^3}{k} (1-e^{-2kr_c\pi}),
\end{eqnarray}
from which we can see that for moderately large values of the
compactification scale $r_c$, the relation between $M_*$ and
$\overline{M}_{P}$ almost does not depend on $r_c$, and it is
completely different from the results in the ADD model. Compared
with the ADD model, the RS model present a different solution to the
gauge Hierarchy problem: the physical mass $m$ of a field on the
brane where our world live on, is related to the fundamental mass
parameter $m_0$ as following
\begin{equation} m=e^{-kr_c\pi}m_0,
\end{equation}
thus the hierarchy problem can be solved if $kr_c\sim 12$.

In the RS model, there also exist KK towers of massive spin-2
gravitons which can interact with the SM fields, and we have the
following 4-dimensional effective Lagrangian\cite{Hewett1,9909255}:
\begin{equation} {\cal L} = - {1\over
\overline{M}_P}T^{\alpha\beta}(x)h^{(0)}_{\alpha\beta}(x)-
{1\over\Lambda_\pi}T^{\alpha\beta}(x)\sum_{n=1}^\infty
h^{(n)}_{\alpha\beta}(x)\, \label{effL} \end{equation}
where
\begin{equation}\label{lambdpi}
\Lambda_\pi=e^{-kr_c\pi}\overline{M}_P=\frac{m_1\overline{M}_P}{x_1k},
\end{equation}
and is at the electroweak scale. Thus the coupling of the massless
graviton $h^{(0)}$ is suppressed by the Planck scale, and the ones
of the massive graviton $h^{(n)}$ by $\Lambda_\pi$, but which is
only TeV. The masses of the $n$th graviton KK excitation modes are
also at the electroweak scale, which are given by
\begin{eqnarray}\label{RSMass} m_n=kx_ne^{-kr_c\pi}=m_1\frac{x_n}{x_1},
\end{eqnarray}
where the $x_n$'s are the $n$th roots of the first order Bessel
function.

From Eqs.(\ref{lambdpi}) and (\ref{RSMass}), the graviton sector of
the RS model is completely determined by the two parameters $m_1$
and $k/\overline{M}_P$. Current
constraints\cite{9909255,0205106,0506158} for the parameters of the
RS model are from the theoretical requirement, the low energy
precise measurement and also the data from Tevatron, from which we
expect $0.01\leq k/\overline{M}_P<0.1$ and $\Lambda_\pi\leq
10$\,TeV.

There are two classes of effects that can be used to probe extra
dimension in the RS model at high energy colliders: real graviton
emission and virtual KK tower exchange. In the RS model, the
lightest massive graviton can have a mass of several hundred GeV,
and may be produced copiously at the LHC. More importantly, it has
much larger couplings to the SM particles than the ones in the ADD
model, thus it may decay into observable particles and hence be
detected. And there have been detailed analysis\cite{0211205}
which demonstrate that using channels $pp\rightarrow
h^{(n)}\rightarrow e^+e^-,\gamma\gamma$..., we can probe the
massive graviton in the RS model with masses up to several TeV.
However, those analysis\cite{0211205} are based on the LO results,
in order to improve the precision of the theoretical predictions,
the higher order QCD effects are necessary. In the ADD model and
the RS model, the NLO QCD corrections to the virtual graviton
production at the LHC have been discussed in Ref.\cite{0506158},
however, the K factors contributed from different parts, the scale
dependence and the PDF uncertainty for above processes needs
further studies. Moreover, they also did not consider the
kinematic distribution of the events, which is very important in
designing the strategy of discovery. In this paper, we study the
transverse momentum distribution of the massive graviton at NLO in
QCD, and all order soft gluon resummation effects on the
distribution to give reasonable predictions.

The arrangement of this paper is as follows. In Sect.~II, we show
the LO results and define the notations. In Sect.~III, we present
the details of the calculations of both the virtual and real parts
of the NLO QCD corrections. In Sect.~IV, we give the transverse
momentum distribution. In Sect.~V, we present the detailed
numerical results for the total cross sections and also the
transverse momentum distribution. Sec.~VI contains a brief
conclusion.

\section{Leading order calculations }
\begin{figure}[ht!]
\vspace{1.0cm} \centerline{\epsfig{file=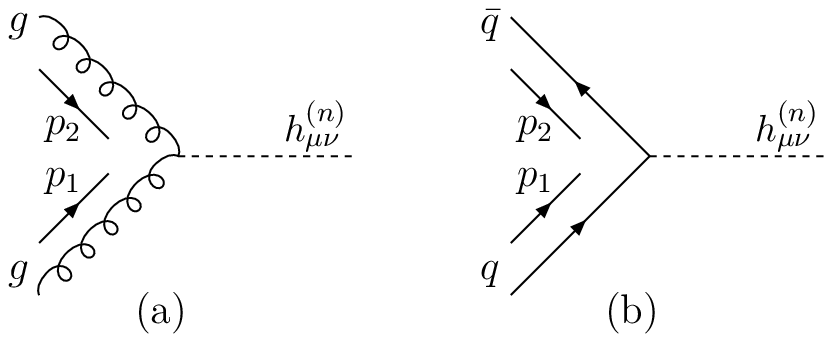, width=240pt}}
\caption[]{Leading order Feynman diagrams for $pp\rightarrow
h^{(n)}_{\mu\nu}$.}\label{9f1}
\end{figure}
The related Feynman diagrams which contribute to the LO amplitude of
the partonic process $g^a_\rho(p_1)g^b_\sigma(p_2)$,
$q_r(p_1)\bar{q}_s(p_2)\rightarrow$ $h^{(n)}_{\mu\nu}$ are shown in
Fig.~\ref{9f1}. The relevant Feynman rules can be read easily from
the ones in the ADD model presented in Ref.\cite{0506158,Hant}, from
which we can get the LO amplitude in $n=4-2\epsilon$ dimensions as
following
\begin{eqnarray}
&&\hspace{-.6cm}M^{(0)}_{gg}=-\frac{i\delta_{ab}\mu_r^{4-n}}{\Lambda_\pi}\times\nonumber
\\&&\hspace{0.3cm}
\bigg[p_1\cdot p_2C_{\mu\nu,\rho\sigma}+D_{\mu\nu,\rho\sigma}
+E_{\mu\nu,\rho\sigma}(p_1,p_2)\bigg]\epsilon^a_\rho(p_1)\epsilon^b_\sigma(p_2)\epsilon^{s*}_{\mu\nu}(p_1+p_2),
\\
&&\hspace{-.6cm}M^{(0)}_{q\bar{q}}=-\frac{i\delta_{rs}\mu_r^{4-n}}{4\Lambda_\pi}\times\nonumber
\\&&\hspace{-0.2cm}\bar{v}(p_2)
\bigg[\gamma_\mu(p_{1\nu}-p_{2\nu})+\gamma_\nu(p_{1\mu}-p_{2\mu})-2\eta_{\mu\nu}(\not{\!p}_1-\not{\!p}_2-2m_q)\bigg]
u(p_1)\epsilon^{s*}_{\mu\nu}(p_1+p_2),
\end{eqnarray}
where $\delta_{ab}$ and $\delta_{rs}$ are color tensors ($a, b$
are the color indices of the initial state gluons, and $r,s$ are
the color indices of the initial state quarks),
$C_{\mu\nu,\rho\sigma}$, $D_{\mu\nu,\rho\sigma}$ and
$E_{\mu\nu,\rho\sigma}(p_1,p_2)$ are the coefficients in the
couplings between the graviton and gluons, which can be found in
Ref.\cite{Hant}, $\mu_r$ is a mass parameter introduced to keep
the couplings dimensionless.

For the polarization sum of the massive graviton, we
have\cite{0506158,Hant}
\begin{eqnarray}
\sum_{s=1}^5\epsilon^{s}_{\mu\nu}(k)\epsilon^{s*}_{\alpha\beta}(k)=P_{\mu\nu\alpha\beta},
\end{eqnarray}
where
 \bea P_{\mu\nu\alpha\beta}  = \frac{1}{2}\left(
\eta_{\mu\alpha}\eta_{\nu\beta} +\eta_{\mu\beta}\eta_{\nu\alpha}
  -\frac{2}{n-1}\eta_{\mu\nu}\eta_{\alpha\beta}\right)+\dots,
 \eea
the dots represent terms proportional to the graviton momentum
$k_{\mu}$, and since $k^{\mu}T_{\mu\nu}=0$, give a vanishing
contribution to the amplitude. For convenience, below we define \bea
\chi\equiv\frac{2}{n-1}=\frac{2}{3-2\epsilon}.\eea

Moreover, in order to avoid introducing external ghost lines while
summing over the gluon helicities, we limit ourselves to the sum
over the physical polarizations of the gluons \cite{beenakker1},
i.e.
\begin{eqnarray}
P_i^{\mu\nu}=\sum_{T} \epsilon_T^\mu(k_i) \epsilon_T^\nu(k_i)=
-g^{\mu\nu} +\frac{n_i^\mu k_i^\nu +k_i^\mu n_i^\nu}{n_i\cdot k_i}
-\frac{n_i^2 k_i^\mu k_i^\nu}{(n_i\cdot k_i)^2},
\end{eqnarray}
where the index $i$ (=1,2) labels the two external gluons, and
$n_i\neq k_i$ is an arbitrary vector. This polarization sum obeys
the transversality relations
\begin{eqnarray}
k_{i\mu}P^{\mu\nu}=P^{\mu\nu} k_{i\nu}=
n_{i\mu}P^{\mu\nu}=P^{\mu\nu} n_{i\nu}=0.
\end{eqnarray}

Thus we can get the relevant partonic cross sections as following:
\begin{eqnarray}
\hat{\sigma}^{(LO)}_{gg}=\frac{1}{2s}2\pi\delta(s-m^2_n)
\overline{|M^{(0)}_{gg}|}^2=\frac{(2-\epsilon)\pi}{32\Lambda^2_\pi}\delta(1-\hat{\tau}),
\end{eqnarray}
\begin{eqnarray}
\hat{\sigma}^{(LO)}_{q\bar{q}}=\frac{1}{2s}2\pi\delta(s-m^2_n)\overline{|M^{(0)}_{q\bar{q}}|}^2=
\frac{(1-\epsilon)\pi}{24\Lambda^2_\pi}\delta(1-\hat{\tau}),
\end{eqnarray}
where $s\equiv(p_1+p_2)^2$ and $\hat{\tau}\equiv m^2_n/s$.

The LO total cross sections at the LHC are obtained by convoluting
the partonic cross sections with the parton distribution functions
(PDFs) $G_{q,\bar{q},g/p}$ in the proton:
\begin{eqnarray}
&&\hspace{-1.5cm}\sigma^{(LO)}\equiv\sigma^{(LO)}_{gg}+\sigma^{(LO)}_{q\bar{q}},\nonumber\vspace{0.7cm}\\
&&\hspace{-1.5cm}\sigma^{(LO)}_{gg}=
\int_{\tau_0}^1dx_1\int_{\tau_0/x_1}^1dx_2\frac{1}{2}\bigg[G_{g/p}(x_1,\mu_f)G_{g/p}(x_2,\mu_f)
+G_{g/p}(x_2,\mu_f)G_{g/p}(x_1,\mu_f)\bigg]\hat{\sigma}^{(LO)}_{gg},\vspace{0.5cm}\\
&&\hspace{-1.5cm}\sigma^{(LO)}_{q\bar{q}}=
\int_{\tau_0}^1dx_1\int_{\tau_0/x_1}^1dx_2\bigg[G_{q/p}(x_1,\mu_f)G_{\bar{q}/p}(x_2,\mu_f)
+G_{q/p}(x_2,\mu_f)G_{\bar{q}/p}(x_1,\mu_f)\bigg]\hat{\sigma}^{(0)}_{q\bar{q}},\end{eqnarray}
where $\tau_0\equiv m^2_n/S_0$, $S_0=(14{\rm TeV})^2$ and $\mu_F$ is
the factorization scale.

\section{Next-to-Leading order calculations}

The NLO QCD corrections consist of the following contributions:
the virtual corrections arising from loop diagrams of colored
particles, the real contributions arising from the radiation of a
real gluon or a massless (anti)quark, and the contributions of
mass factorization. In the following, we will calculate these
contributions separately. We use dimensional regularization
(DREG)\cite{DREG} in $d=4-2\epsilon$ dimensions to regulate the
ultraviolet (UV) and infrared (IR) divergences.

For the partonic cross section, the total virtual corrections can
be written as
\begin{eqnarray}
\hat{\sigma}^{V}_{gg,q\bar{q}}=\hat{\sigma}^{unren}_{gg,q\bar{q}}+\hat{\sigma}^{con}_{gg,q\bar{q}},\label{9rn}
\end{eqnarray}
where the first part in the right hand contains the radiative
corrections from the one-loop vertex diagrams, and the second part
is the contributions from the counterterms involving only the the
wavefunction renormalization constant for the external fields.

The ${\cal O} (\alpha_s)$ virtual corrections to the partonic cross
section can be expressed as
\begin{eqnarray}
&&\hat{\sigma}^{V}_{gg}=2C_\epsilon\frac{g^2_s(2-\epsilon)}{32\pi\Lambda^2_\pi}\delta(1-\hat{\tau})\nonumber\\
&&\hspace{2.cm}\times
\bigg(\frac{-3}{8}\frac{1}{\epsilon^2_{IR}}-\frac{33}{48}\frac{1}{\epsilon_{IR}}+\frac{n_f}{24}\frac{1}{\epsilon_{IR}}
+\frac{1}{8}\pi^2-\frac{203}{96}+\frac{35n_f}{288}\bigg),\\
&&\hat{\sigma}^{V}_{q\bar{q}}=2C_\epsilon\frac{g^2_s(1-\epsilon)}{24\pi\Lambda^2_\pi}\delta(1-\hat{\tau})\times
\bigg(\frac{-1}{6}\frac{1}{\epsilon^2_{IR}}-\frac{1}{3}\frac{1}{\epsilon_{IR}}
+\frac{1}{18}\pi^2-\frac{5}{6}\bigg), \end{eqnarray} with
$C_\epsilon\equiv\frac{\Gamma(1-\epsilon)}{\Gamma(1-2\epsilon)}\bigg(\frac{4\pi\mu^2_r}
{m^2_n}\bigg)^\epsilon$, which are UV finite, but still contain the
IR divergences. Here, the IR divergences include the soft
divergences and the collinear divergences. The soft divergences are
canceled after adding the real emission corrections, and the
remaining collinear divergences can be absorbed into the
redefinition of PDF \cite{altarelli}, which will be discussed in the
following subsections.

\begin{figure}[ht!]
\vspace{1.0cm} \centerline{\epsfig{file=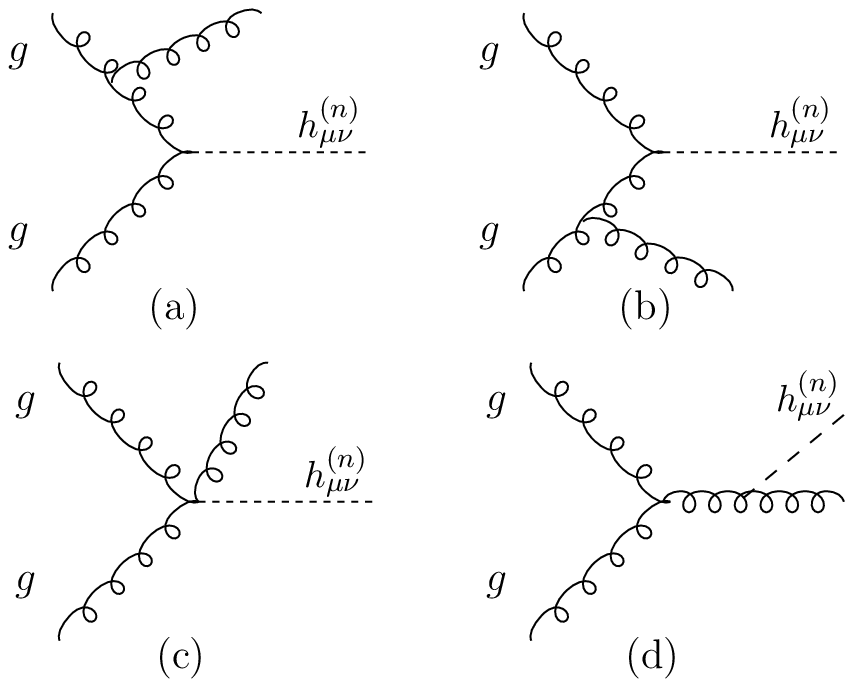, width=240pt}}
\caption[]{Feynman diagrams of real gluon emission sub-processes
$gg\rightarrow h^{(n)}_{\mu\nu}+g$.}\label{9f4}
\end{figure}

\begin{figure}[ht!]
\vspace{1.0cm} \centerline{\epsfig{file=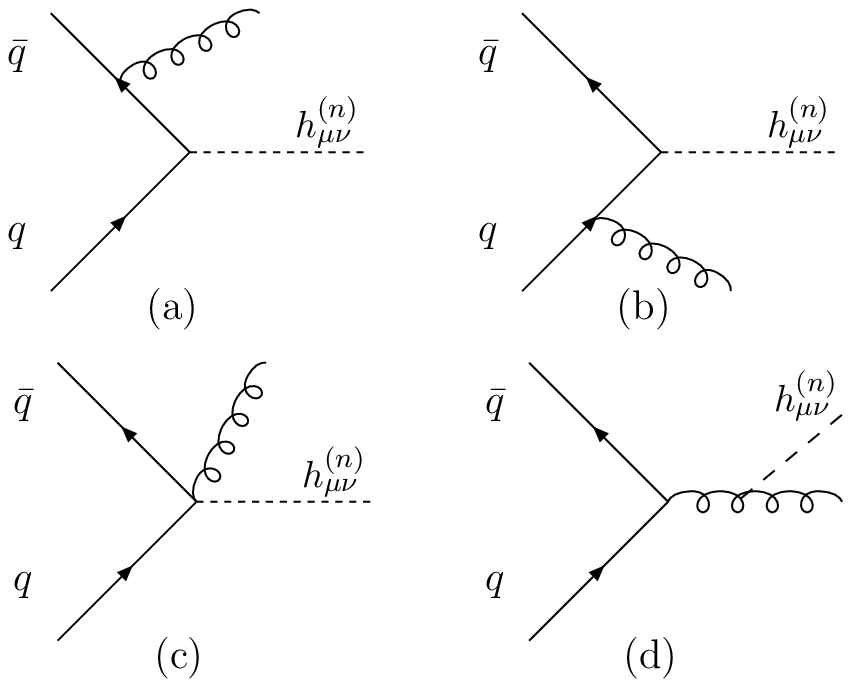, width=240pt}}
\caption[]{Feynman diagrams of real gluon emission sub-processes
$q\bar{q}\rightarrow h^{(n)}_{\mu\nu}+g$.}\label{9f5}
\end{figure}

\begin{figure}[ht!]
\vspace{1.0cm} \centerline{\epsfig{file=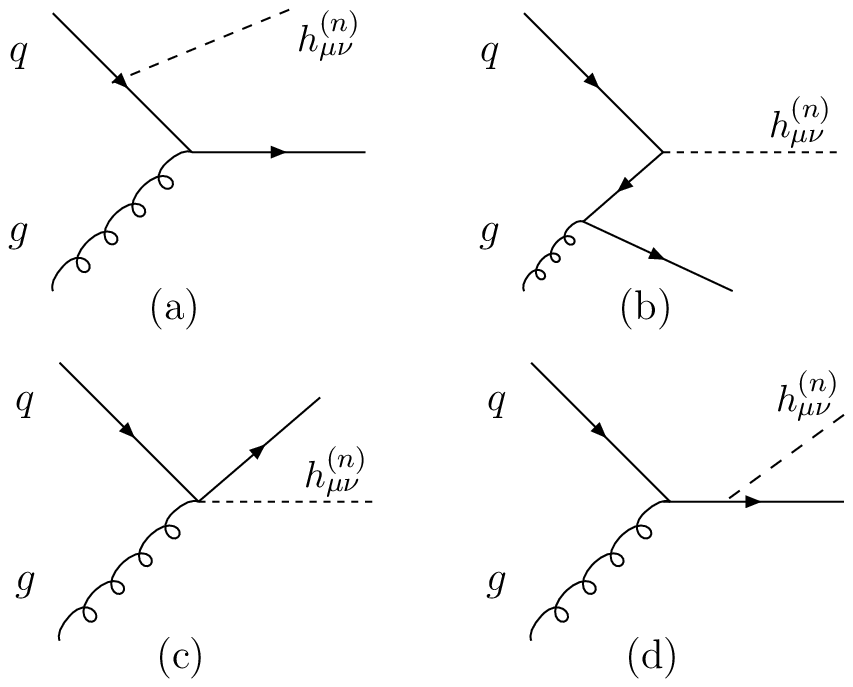, width=240pt}}
\caption[]{Feynman diagrams of real quark emission sub-processes
$gq\rightarrow h^{(n)}_{\mu\nu}+q$.}\label{9f6}
\end{figure}

The real corrections consist of the contributions from the radiation
of a real gluon or a massless (anti)quark. The Feynman diagrams for
the real gluon emission sub-process $gg,q\bar{q}\rightarrow
h^{(n)}_{\mu\nu}$ are shown in Figs.~\ref{9f4} and Fig.~\ref{9f5},
respectively. The Feynman diagrams for massless (anti)quark emission
(the diagrams for the antiquark emission sub-processes are similar
and omitted here) are shown in Fig.~\ref{9f6}

For the real gluon emission sub-processes $g(p_1)g(p_2),\,\,
q(p_1)\bar{q}(p_2)\rightarrow g(p_3)h^{(n)}_{\mu\nu}$, the partonic
cross sections is£º
\begin{eqnarray}
\hat{\sigma}_{gg,q\bar{q}}^{real}=\frac{1}{2s}\int
{\overline{|M^{real}_{gg,q\bar{q}}|}}^2d\Gamma_2,
\end{eqnarray}
with
\begin{eqnarray}\label{rg1}&&\hspace{-0.5cm}{\overline{|M^{real}_{gg}|}}^2=\frac{3\times
8}{8\times
8}\frac{1}{4(1-\epsilon)^2}\frac{4g_s^2}{\Lambda^2_\pi}\frac{1}{tu}\times
\bigg\{ \epsilon
t^2u[26-9(4-2\epsilon)-2(-5+2(4-2\epsilon))\epsilon\chi]\nonumber\\
&&+\epsilon^2\chi tu^2[26-9(4-2\epsilon)-2(-5+2(4-2\epsilon))]
-\frac{(1-\epsilon)}{4}s^3[16-6(4-2\epsilon)+4\epsilon^2\chi]\nonumber\\
&&-\frac{(1-\epsilon)}{2}t^3[16-6(4-2\epsilon)+4\epsilon^2\chi]
-\frac{(1-\epsilon)}{2}s^2(t+u)[16-
6(4-2\epsilon)+4\epsilon^2\chi]\nonumber\\
&&-\frac{(1-\epsilon)}{2}u^3[16-6(4-2\epsilon)+4\epsilon^2\chi]
-\frac{(1-\epsilon)}{4s}(t^2+tu+u^2)^2[16-6(4-2\epsilon)+4\epsilon^2\chi]\nonumber\\
&&+\frac{s}{2}[2\epsilon tu(26-9(4-2\epsilon)-6\epsilon\chi)
-\frac{3}{2}(1-\epsilon)(t^2+u^2)[16-6(4-2\epsilon)+4\epsilon^2\chi]\bigg\},
\end{eqnarray}
\begin{eqnarray}\label{rg2}{\overline{|M^{real}_{q\bar{q}}|}}^2&=&\frac{4}{3\times
3}\frac{1}{2\times
2}\frac{-g_s^2}{\Lambda^2_\pi}\frac{1}{stu}\times
 \bigg\{2(\epsilon-1)s^4+4(\epsilon-1)(t+u)s^3\nonumber\\
&&
-(\epsilon-3)[(\epsilon-1)t^2+2(3\epsilon-2)ut+(\epsilon-1)u^2]s^2
\nonumber\\
&&
-(t+u)[(\epsilon-1)^2t^2+(6\epsilon(\epsilon-3)+8)ut+(\epsilon-1)^2u^2]s\nonumber\\
&& -2tu[(\epsilon-1)t^2+2\epsilon
ut+(\epsilon-1)u^2][\epsilon(\epsilon\chi+3)-2]\bigg\},
\end{eqnarray}
where ${\overline{|M^{real}_{gg,q\bar{q}}|}}^2$ is the squared
matrix of the real gluon emission sub-processes, in which the
colors and spins of the outgoing particles have been summed, and
the colors and spins of the incoming ones have been averaged over,
and the final state 2-body phase space is
\begin{eqnarray}
d\Gamma_2=\frac{1}{8\pi\Gamma(1-\epsilon)}\bigg(\frac{4\pi\mu^2_r}{m^2_n}\bigg)^\epsilon
(\hat{\tau})^\epsilon(1-\hat{\tau})^\epsilon\times v^{-\epsilon}
(1-v)^{-\epsilon}dv,
\end{eqnarray}
where \begin{eqnarray} &&v\equiv \frac{1}{2}(1+\cos\theta),\\
&&t\equiv(p_1-p_3)^2=-s(1-\hat{\tau})(1-v),\\
&&u\equiv(p_2-p_3)^2=-s(1-\hat{\tau})v,
\end{eqnarray}
and $\theta$ is the the angle between the incoming gluon and the
outgoing gluon.

Combining the contributions of the virtual corrections and the
real gluon emission, we still have the collinear divergences,
which can be absorbed into the redefinition of the PDF at NLO, in
general called mass factorization \cite{altarelli}. This procedure
in practice means that first we convolute the partonic cross
section with the bare PDF $G_{\alpha/p}(x)$, and then rewrite
$G_{\alpha/p}(x)$ in terms of the renormalized PDF
$G_{\alpha/p}(x,\mu_f)$in the numerical calculations. In the
$\overline{\rm MS}$ scheme, the scale dependent PDF
$G_{\alpha/p}(x,\mu_f)$ is given by\cite{cutoff}
\begin{eqnarray}
G_{\alpha/p}(x,\mu_f)= G_{\alpha/p}(x)+
\sum_{\beta}(-\frac{1}{\epsilon})\bigg [\frac{\alpha_s}{2\pi}
\frac{\Gamma(1 -\epsilon)}{\Gamma(1 -2\epsilon)} \bigg(\frac{4\pi
\mu_r^2}{\mu_f^2}\bigg)^\epsilon\bigg]  \int_x^1 \frac{dz}{z}
P_{\alpha\beta} (z) G_{\beta/p}(x/z),
\end{eqnarray}
where $P_{\alpha\beta}(z)$ are the leading order Altarelli-Parisi
splitting functions \cite{alpha}.

Here, for the real gluon emission sub-processes $gg\rightarrow
h^{(n)}_{\mu\nu}$ and $q\bar{q}\rightarrow h^{(n)}_{\mu\nu}$, we
first consider only the contributions from $p_{gg}$ and $p_{qq}$,
and we can get the relevant counterterm arising from the PDF
redefinition as following:
\begin{eqnarray}
\hspace{-1.4cm}\delta\hat{\sigma}_{gg}
=2\times\frac{\alpha_s}{2\pi}C'_\epsilon\frac{(2-\epsilon)}{32\Lambda^2_\pi}zP^{(0)}_{gg}(z)
,\end{eqnarray}
\begin{eqnarray}
\hspace{-1.4cm}\delta\hat{\sigma}_{q\bar{q}}
=2\times\frac{\alpha_s}{2\pi}C'_\epsilon\frac{(1-\epsilon)}{24\Lambda^2_\pi}zP^{(0)}_{qq}(z)
,\end{eqnarray} where
$C'_\epsilon\equiv\frac{\Gamma(1-\epsilon)}{\Gamma(1-2\epsilon)}\bigg(\frac{4\pi\mu^2_r}
{\mu^2_f}\bigg)^\epsilon$ and $z\equiv
m^2_n/(x_1x_2S)=\hat{\tau}$.

Summing up the virtual, real emission and PDF redefinition
contributions, we have the IR finite results
\begin{eqnarray}\label{77gg}
&&\hspace{-1.1cm}\hat{\sigma}_{gg}^{NLO}=\hat{\sigma}_{gg}^{real}+\hat{\sigma}_{gg}^{V}+\delta\hat{\sigma}_{gg}
=\frac{(2-\epsilon)\alpha_s}{32\Lambda^2_\pi}C_\epsilon\frac{m^2_n}{s}\times
\nonumber\\
&&\bigg\{
6\ln\bigg(\frac{m^2_n}{\mu^2_f}\bigg)[\frac{\hat{\tau}}{(1-\hat{\tau})_+}+\frac{1-\hat{\tau}}{\hat{\tau}}
+\hat{\tau}(1-\hat{\tau})]+\ln\bigg(\frac{m^2_n}{\mu^2_f}\bigg)(\frac{11}{2}-\frac{n_f}{3})\delta(1-\hat{\tau})
\nonumber\\
&&
+(\pi^2-\frac{203}{12}+\frac{35n_f}{36})\delta(1-\hat{\tau})+12\bigg(\frac{\ln(1-\hat{\tau})}{1-\hat{\tau}}\bigg)_+
\nonumber\\
&&+6[-1+\frac{1-\hat{\tau}}{\hat{\tau}}
+\hat{\tau}(1-\hat{\tau})]\ln\bigg(\frac{(1-\hat{\tau})^2}{\hat{\tau}}\bigg)-6\frac{\ln\hat{\tau}}{1-\hat{\tau}}
-\frac{3}{2}-\frac{11}{2\hat{\tau}}+\frac{3}{2}\hat{\tau}+\frac{11\hat{\tau}^2}{2}\bigg\},
\\
&&\hspace{-1.1cm}\hat{\sigma}_{q\bar{q}}^{NLO}=\hat{\sigma}_{q\bar{q}}^{real}+\hat{\sigma}_{q\bar{q}}^{V}
+\delta\hat{\sigma}_{q\bar{q}}
=\frac{(1-\epsilon)\alpha_s}{24\Lambda^2_\pi}C_\epsilon\frac{m^2_n}{s}\nonumber\\
&&\times\bigg\{
\frac{4}{3}\ln\bigg(\frac{m^2_n}{\mu^2_f}\bigg)[\frac{1+\hat{\tau}^2}{(1-\hat{\tau})_+}+\frac{3}{2}\delta(1-\hat{\tau})]
+\frac{4}{3}(-5+\frac{\pi^2}{3})\delta(1-\hat{\tau})
+\frac{16}{3}\bigg(\frac{\ln(1-\hat{\tau})}{1-\hat{\tau}}\bigg)_+\nonumber\\
&&-\frac{4}{3}(1+\hat{\tau})\ln\frac{(1-\hat{\tau})^2}{\hat{\tau}}
-\frac{8}{3}\frac{\ln\hat{\tau}}{1-\hat{\tau}}+\frac{16}{9\hat{\tau}}
-\frac{16\hat{\tau}^2}{9}\bigg\}.\nonumber
\end{eqnarray}

For the real (anti-)quark emission sub-processes
$gq(\bar{q})\rightarrow q(\bar{q})h^{(n)}_{\mu\nu}$, the relevant
results can can be got in the similar way as above. First, the
partonic cross sections is£º
\begin{eqnarray}
\hat{\sigma}_{gq,g\bar{q}}^{real}=\frac{1}{2s}\int
{\overline{|M^{real}_{gq,g\bar{q}}|}}^2d\Gamma_2,
\end{eqnarray}
with
\begin{eqnarray}\label{rg3}
 &&{\overline{|M^{real}_{gq}|}}^2=-\frac{1}{1-\epsilon}{\overline{|M^{real}_{q\bar{q}}|}}^2
 (s\leftrightarrow
 t),\\
  &&{\overline{|M^{real}_{g\bar{q}}|}}^2=
 -\frac{1}{1-\epsilon}{\overline{|M^{real}_{q\bar{q}}|}}^2(s\leftrightarrow
 u),
\end{eqnarray}
where ${\overline{|M^{real}_{gq,g\bar{q}}|}}^2$ is the squared
matrix of the real (anti-)quark emission sub-processes, in which
the colors and spins of the outgoing particles have been summed,
and the colors and spins of the incoming ones have been averaged
over.

Secondly, the relevant counterterm arising from the PDF redefinition
for real (anti-)quark emission sub-processes are shown as following,
where we consider only the contributions from $p_{gq}$ and $p_{qg}$:
\begin{eqnarray} \hspace{-1.4cm}\delta\hat{\sigma}_{gq}
=\frac{\alpha_s}{2\pi}C'_\epsilon\frac{(1-\epsilon)}{24\Lambda^2_\pi}zP^{(0)}_{qg}(z)
+2\times\frac{1}{2}\frac{\alpha_s}{2\pi}C'_\epsilon\frac{(2-\epsilon)}{32\Lambda^2_\pi}zP^{(0)}_{gq}(z).
\end{eqnarray}

Summing up the virtual, real emission and PDF redefinition
contributions, again we get the IR finite results for real
(anti-)quark emission sub-processes,

\begin{eqnarray}\label{77gq}
&&\hspace{-1.3cm}\hat{\sigma}_{gq}^{NLO}(=\hat{\sigma}_{g\bar{q}}^{NLO})
=\hat{\sigma}_{gq}^{real} +\delta\hat{\sigma}_{gq}
=\frac{\alpha_s}{96\Lambda^2_\pi}C_\epsilon\frac{m^2_n}{s}\nonumber\\
&&\hspace{-0.9cm}\times\bigg\{[4\frac{1+(1-{\hat\tau})^2}{\hat{\tau}}+((1-\hat{\tau})^2+\hat{\tau}^2)]\ln\bigg(\frac{m^2_n}{\mu^2_f}\bigg)
\nonumber\\
&&\hspace{-0.9cm}+[4\frac{1+(1-\hat{\tau})^2}{\hat{\tau}}+((1-\hat{\tau})^2+\hat{\tau}^2)]\ln\bigg(\frac{(1-\hat{\tau})^2}{\hat{\tau}}\bigg)
+\frac{9}{2}-\frac{6}{\hat{\tau}}+9\hat{\tau}-\frac{7\hat{\tau}^2}{2}\bigg\}.
\end{eqnarray}

The NLO total cross section for $pp\rightarrow h^{(n)}$ in the
$\overline{MS}$ factorization scheme is obtained by summing up the
Born, virtual, real emission and PDF redefinition contributions.
In terms of the above notations, we have
\begin{eqnarray}
&&\hspace{-0.4cm}\sigma^{(NLO)}=\sigma^{(LO)}_{gg}+\sigma^{(LO)}_{q\bar{q}}
+\sigma^{(NLO)}_{gg}+\sigma^{(NLO)}_{q\bar{q}}
+\sigma^{(NLO)}_{gq}+\sigma^{(NLO)}_{g\bar{q}}\nonumber \\ &&
\hspace{0.9cm}=\sigma^{(LO)}_{gg}+\sigma^{(LO)}_{q\bar{q}}\nonumber \\
&&\hspace{0.7cm}
+\int_{\tau_0}^1dx_1\int_{\tau_0/x_1}^1dx_2\frac{1}{2}\bigg[G_{g/p}(x_1,\mu_f)G_{g/p}(x_2,\mu_f)
+G_{g/p}(x_2,\mu_f)G_{g/p}(x_1,\mu_f)\bigg]\hat{\sigma}^{(NLO)}_{gg}\nonumber
\\ &&\hspace{0.7cm}+
\int_{\tau_0}^1dx_1\int_{\tau_0/x_1}^1dx_2\bigg[G_{q/p}(x_1,\mu_f)G_{\bar{q}/p}(x_2,\mu_f)
+G_{q/p}(x_2,\mu_f)G_{\bar{q}/p}(x_1,\mu_f)\bigg]\hat{\sigma}^{(NLO)}_{q\bar{q}}
\nonumber \\ &&\hspace{0.7cm}
+\int_{\tau_0}^1dx_1\int_{\tau_0/x_1}^1dx_2\bigg[G_{q/p}(x_1,\mu_f)G_{g/p}(x_2,\mu_f)
+G_{q/p}(x_2,\mu_f)G_{g/p}(x_1,\mu_f)\bigg]\hat{\sigma}^{(NLO)}_{gq}
\nonumber \\ &&\hspace{0.7cm}
+\int_{\tau_0}^1dx_1\int_{\tau_0/x_1}^1dx_2\bigg[G_{g/p}(x_1,\mu_f)G_{\bar{q}/p}(x_2,\mu_f)
+G_{g/p}(x_2,\mu_f)G_{\bar{q}/p}(x_1,\mu_f)\bigg]\hat{\sigma}^{(NLO)}_{g\bar{q}}.
\end{eqnarray}

Finally, we note that our NLO results in Eqs.(\ref{77gg}) and
(\ref{77gq}) are the same as the ones in Eq.(3.38) of the second
paper in Ref.\cite{0506158}, except the differences by overall
factors.

\section{Transverse momentum distribution}

In this section we investigate the transverse momentum distribution
of the massive graviton. At LO the graviton is kept at zero $q_T$
due to momentum conservation and the distribution is proportional to
$\delta^2(\vec{q}_T)$. Thus the LO distribution at non-zero $q_T$
belongs to $\mathcal{O}(\alpha_s)$, where momentum conservation is
retained by the additional parton emitted. The distribution can be
obtained from the squared amplitudes of the real emission processes,
i.e. Eqs.~(\ref{rg1}), (\ref{rg2}) and (\ref{rg3}). However, the
corresponding fixed order result of the transverse momentum
distribution is only valid when $q_T$ is not too small compared with
the mass of the massive graviton $m_n$. If $q_T \ll m_n$, the
corresponding parton emitted would be either soft or collinear to
one of the initial partons. Thus, large logarithms like
$\ln(m_n^2/q_T^2)$ will appear and will dominate over the cross
section for sufficiently small $q_T$. In general, there should be
double logarithms for each gluon attached to the initial quarks due
to the overlap of soft region and collinear region. As a result, the
perturbative expansion would be controlled by
$\alpha_s\ln^2(m_n^2/q_T^2)$ rather than $\alpha_s$. The convergence
of the perturbation series will be spoiled if
$\alpha_s\ln^2(m_n^2/q_T^2)$ approaches unity. In order to make use
of the perturbation theory with the existence of large logarithms at
each order, one must reorganize the perturbative expansion to resum
the large terms. In this paper, we use the Collins-Soper-Sterman
(CSS) resummation formalism\cite{Nucl.Phys.B250.199} to calculate
all order soft gluon effects on the transverse momentum
distribution.

In the CSS formalism, the differential cross section we are
considering can be written as
\begin{equation}
  \label{eq:total00}
  \frac{d\sigma}{dq_T^2dy}({\rm total}) = \frac{d\sigma}{dq_T^2dy}
  ({\rm resum}) + Y(q_T,m,x_1^0,x_2^0),
\end{equation}
where \begin{eqnarray}
  \label{eq:total11}
  \frac{d\sigma}{dq_T^2dy}
  ({\rm resum}) = \sum_{\alpha,\beta}\frac{d\sigma_{\alpha\beta}}{dq_T^2dy}
  ({\rm resum})\\ \label{eq:total12}
  Y(q_T,m,x_1^0,x_2^0)=\sum_{ab}Y_{ab}(q_T,m,x_1^0,x_2^0),
  \end{eqnarray}
  and the resummed part can be expressed as an inverse Fourier transformation
  \begin{eqnarray}
  \label{eq:resum}
  \frac{d\sigma_{\alpha\beta}}{dq_T^2dy} ({\rm resum}) &=&
  \frac{1}{2} \sigma^0_{\alpha\beta} \frac{1}{2\pi} \int d^2\vec{b} \exp \left( i\vec{b} \cdot
    \vec{q}_T \right) W_{\alpha\beta}(b,m,x_1^0,x_2^0)\nonumber
 \\
  &&= \sum_{\alpha,\beta} \frac{1}{2} \sigma^0_{\alpha\beta} \int_0^\infty b db J_0(b q_T)
  W_{\alpha\beta}(b,m,x_1^0,x_2^0),
\end{eqnarray}
with
\begin{eqnarray}\label{opos}
 \hspace{-0.8cm} W_{\alpha\beta}(b,m,x_1^0,x_2^0) &=& \tilde{f}_{\alpha/A}(x_1^0,C_3/b)
  \tilde{f}_{\beta/B}(x_2^0,C_3/b) \nonumber
  \\
 \hspace{-0.8cm}  \qquad \times &\exp &\left\{ - \int_{C_1^2/b^2}^{C_2^2m^2}
    \frac{d\bar{\mu}^2}{\bar{\mu}^2} \left[ \ln\frac{C_2^2m^2}{\bar{\mu}^2}
      A(\alpha_s(\bar{\mu})) + B(\alpha_s(\bar{\mu})) \right]
      \right\},
\end{eqnarray}
where $\alpha\beta=gg,q\bar{q}$, $ab=gg,q\bar{q},q(\bar{q})g$,
$\vec{b}$ is the impact parameter conjugating to $\vec{q}_T$, $J_0$
is zero order Bessel function of the first kind, and
$x^0_1=e^ym_n/\sqrt{s}$, $x^0_2=e^{-y}m_n/\sqrt{s}$. Here $C_i
(i=1,2,3)$ are constants of order 1 which are by convention
\cite{Nucl.Phys.B250.199} chosen to be
\begin{equation}
  C_1 = C_3 = 2 e^{-\gamma_E} \equiv b_0, \, C_2 = 1,
\end{equation}
and $\tilde{f}$ is the convolution of the PDFs and the coefficient
functions $C$
\begin{equation}
  \tilde{f}_{\alpha/h}(x,\mu) = \sum_{\gamma} \int_x^1 \frac{dz}{z}
  C_{\alpha\gamma}(z,\alpha_s(\mu)) f_{\gamma/h}(x,\mu),
\end{equation}
and the coefficients $A$, $B$ and $C$ can be expanded to series in
$\alpha_s$
\begin{eqnarray}
  A(\alpha_s) &=& \sum_{n=1}^\infty A^{(n)} \left( \frac{\alpha_s}{\pi}
  \right)^n,
  \\
  B(\alpha_s) &=& \sum_{n=1}^\infty B^{(n)} \left( \frac{\alpha_s}{\pi} \right)^n,
  \\
  C_{\alpha\beta}(z,\alpha_s) &=& \sum_{n=0}^\infty C_{\alpha\beta}^{(n)}(z) \left(
    \frac{\alpha_s}{\pi} \right)^n,
\end{eqnarray}
and they can be calculated order by order in perturbative theory.
In our case, since the massive graviton is colorless, thus the
lowest order coefficients is the same as the ones in the case of
$gg\rightarrow H^0$\cite{hhgres} and Drell-Yan\cite{dygres}. For
the $gg$ channel, we have
\begin{eqnarray}
 && A^{(1)} = 2N_c = 6 , \qquad B^{(1)} = -2\beta_0 =(33-2n_f)/6,\nonumber
  \\
 && C_{\alpha\beta}^{(0)}(z) = \delta_{\alpha\beta}
  \delta(1-z),
\end{eqnarray}
and for the $q\bar{q}$ channel, we have
\begin{eqnarray}
  && A^{(1)} = C_F = \frac{4}{3} , \qquad B^{(1)} = - \frac{3}{2} C_F = -2 ,
  \\
  && C_{\alpha\beta}^{(0)}(z) = \delta_{\alpha\beta} \delta(1-z).
\end{eqnarray}
With these coefficients, we can actually sum up all terms like
$\alpha_s^nL^{2n-1}$ and $\alpha_s^nL^{2n-2}$.

However, the resummed part is still not able to be calculated
perturbatively. The reason is that in Eq.~(\ref{eq:resum}), the
integral over the impact parameter $b$ extends to infinity, while
the integrand involves the strong coupling constant $\alpha_s$ and
the PDFs at scale $b_0/b$, where they are not well defined if $b$ is
large enough so that $b_0/b$ enters non-perturbative region.
Collins, Soper and Sterman, in their original paper
\cite{Nucl.Phys.B250.199}, suggested that one can use a cut-off
$b_{\text{max}}$ and regard the effects from $b>b_{\text{max}}$ as
non-perturbative input. Practically, they replaced $W(b)$ in
Eq.~(\ref{eq:resum}) by
\begin{equation}
  \widetilde{W}(b) = W(b_*) F_{\text{NP}}(b),
\end{equation}
where
\begin{equation}
  b_* = \frac{b}{\sqrt{1+(b/b_{\text{max}})^2}},
\end{equation}
and $F_{\text{NP}}(b)$ parameterizes the non-perturbative effects.
Since $b_*$ never exceeds $b_{\text{max}}$, $W(b_*)$ can be
calculated perturbatively, and the theoretical uncertainty mainly
relies on the function $F_{\text{NP}}$. Recently, Landry, Brock,
Nadolsky and Yuan (BLNY) \cite{Phys.Rev.D67.073016} proposed the
form
\begin{equation}
  F_{{\rm NP}} = \exp \left\{ - b^2 \left[ g_1 + g_2 \ln\frac{m}{2Q_0} +
      g_1g_3\ln(100x_1^0x_2^0) \right] \right\},\nonumber
\end{equation}
They take $b_{\text{max}} = 0.5 \text{GeV}^{-1}$, $Q_0 = 1.6
\text{GeV}$ and the parameters $g_i (i=1,2,3)$ are fitted to the
available Drell-Yan data, which are given by
\begin{equation}
  g_1 = 0.21, \qquad g_2 = 0.68, \qquad g_3 = -0.60.
\end{equation}

The another term in Eq.~(\ref{eq:total11}), the $Y$ term, is the
remaining contributions which are not resummed. Since it contains
no large logarithms, it can be reliably computed by fixed order
truncation of the perturbative series
\begin{equation}\label{asyys}
  Y_{ab} = \frac{d\sigma_{ab}}{dq_T^2dy} ({\rm pert}) -
  \frac{d\sigma_{ab}}{dq_T^2dy} ({\rm asym}),
\end{equation}
where the first term in the right hand is the fixed-order
perturbative results, and the second term is the asymptotic part
of the differential cross section, defined as the terms which are
at least as singular as $1/q_T^2$ when $q_T \to 0$, which can be
got by expanding the resummed part, i.e. Eq.(\ref{eq:resum}). In
our case, we have
\begin{eqnarray}
  \label{eq:asym1}
&& \frac{d\sigma_{gg}}{dq^2_Tdy}({\rm asym}) = \frac{1}{2}
\sigma^0_{gg}
  \frac{\alpha_s}{2\pi} \frac{\tau_0}{q^2_T} \biggl\{ f_{g/P}(x_1^0,\mu_f)
    f_{g/P}(x_2^0,\mu_f) \left( 2N_c \ln\frac{m^2}{q_T^2} - 2\beta_0 \right) \biggr.
  \nonumber \\
  &&\qquad + \biggl. (P_{gg}\circ{f})_{g/P}(x_1^0,\mu_f) f_{g/P}(x_2^0,\mu_f) +
      f_{g/P}(x_1^0,\mu_f) (P_{gg}\circ{f})_{g/P}(x_2^0,\mu_f)\nonumber \\
  &&\qquad +(x_1^0\leftrightarrow x_2^0) \biggr\},
\end{eqnarray}
\begin{eqnarray}
  \label{eq:asym2}
&& \frac{d\sigma_{q\bar{q}}}{dq^2_Tdy}({\rm asym}) =
\sigma^0_{q\bar{q}}
  \frac{\alpha_s}{2\pi} \frac{\tau_0}{q^2_T} \biggl\{ f_{q/P}(x_1^0,\mu_f)
    f_{\bar{q}/P}(x_2^0,\mu_f) \left( 2C_F \ln\frac{m^2}{q_T^2} - 3C_F \right) \biggr.
  \nonumber \\
  &&\qquad + \biggl. (P_{qq}\circ{f})_{\bar{q}/P}(x_2^0,\mu_f) f_{q/P}(x_1^0,\mu_f) +
      f_{\bar{q}/P}(x_2^0,\mu_f) (P_{qq}\circ{f})_{q/P}(x_1^0,\mu_f)\nonumber \\
  &&\qquad +(x_1^0\leftrightarrow x_2^0) \biggr\},
\end{eqnarray}
\begin{eqnarray}
  \label{eq:asym3}
&& \frac{d\sigma_{qg}}{dq^2_Tdy}({\rm asym}) =  \sigma^0_{q\bar{q}}
  \frac{\alpha_s}{2\pi} \frac{\tau_0}{q^2_T} \bigg[
      f_{q/P}(x_1^0,\mu_f) (P_{qg}\circ{f})_{g/P}(x_2^0,\mu_f)\bigg]\nonumber \\
   &&\hspace{1.4cm}+\frac{1}{2}\sigma^0_{gg}
  \frac{\alpha_s}{2\pi} \frac{\tau_0}{q^2_T} \bigg[
      f_{g/P}(x_1^0,\mu_f) (P_{gq}\circ{f})_{q/P}(x_2^0,\mu_f)\bigg]+(x_1^0\leftrightarrow x_2^0),
\end{eqnarray}
where $\sigma^0_{gg}\equiv\frac{\pi}{16\Lambda^2_\pi}$ and
$\sigma^0_{q\bar{q}}\equiv\frac{\pi}{24\Lambda^2_\pi}$. The result
of $\frac{d\sigma_{\bar{q}g}}{dq^2_Tdy}(asym)$ is similar to the
one of $\frac{d\sigma_{qg}}{dq^2_Tdy}(asym)$, and thus omitted
here.

\section{Numerical Results}

As mentioned in Section I, there are two additional free inputs in
the RS model: $m_1$ and $k/\overline{M}_P$, which have the following
constraints: $0.01\leq k/\overline{M}_P<0.1$ and
$\Lambda_\pi=\frac{m_1\overline{M}_P}{x_1k}\leq 10$\,TeV. In our
numerical calculations, for convenience, we choose the input
parameters as $\Lambda_\pi$ and $m_1$. For $\Lambda_\pi=4\,(8)TeV$,
from current constraints, we have $150\,{\rm GeV}<m_1<1.5\,{\rm
TeV}$ ($300\,{\rm GeV}<m_1<3\,{\rm TeV}$).

Moreover, for the NLO total cross sections $\sigma^{(NLO)}$ and the
contributions from different parts (including $\sigma_{gg}^{(NLO)}$,
$\sigma_{q\bar{q}}^{(NLO)}$ and
$\sigma_{gq}^{(NLO)}+\sigma_{g\bar{q}}^{(NLO)}$), the NLO
($\overline{\text{MS}}$) PDFs \cite{CTEQ} is used throughout this
paper. For the LO results, we define two cross sections as
following:
\begin{eqnarray}
  \sigma^{\mathrm{(LO1)}}: &\quad \text{LO partonic cross section convoluted with NLO
    ($\overline{\text{MS}}$) PDFs};
  \\
  \sigma^{\mathrm{(LO2)}}: &\quad \text{LO partonic cross section convoluted with LO PDFs},
  \end{eqnarray}
and correspondingly two $K$ factors:
\begin{equation}
  K_1 = \frac{\sigma^{\mathrm{(NLO)}}}{\sigma^{\mathrm{(LO1)}}}, \qquad
  K_2 = \frac{\sigma^{\mathrm{(NLO)}}}{\sigma^{\mathrm{(LO2)}}}.
\end{equation}
As the above definitions, $K_1$ measures only the size of the NLO
QCD corrections to the cross sections, while $K_2$ accounts for the
effects of changing parton distribution functions additionally. As
for the renormalization and factorization scales, we always choose
$\mu_r=\mu_f=m_n$, unless otherwise specified.

\begin{figure}[ht!] \vspace{1.0cm}
\centerline{\epsfig{file=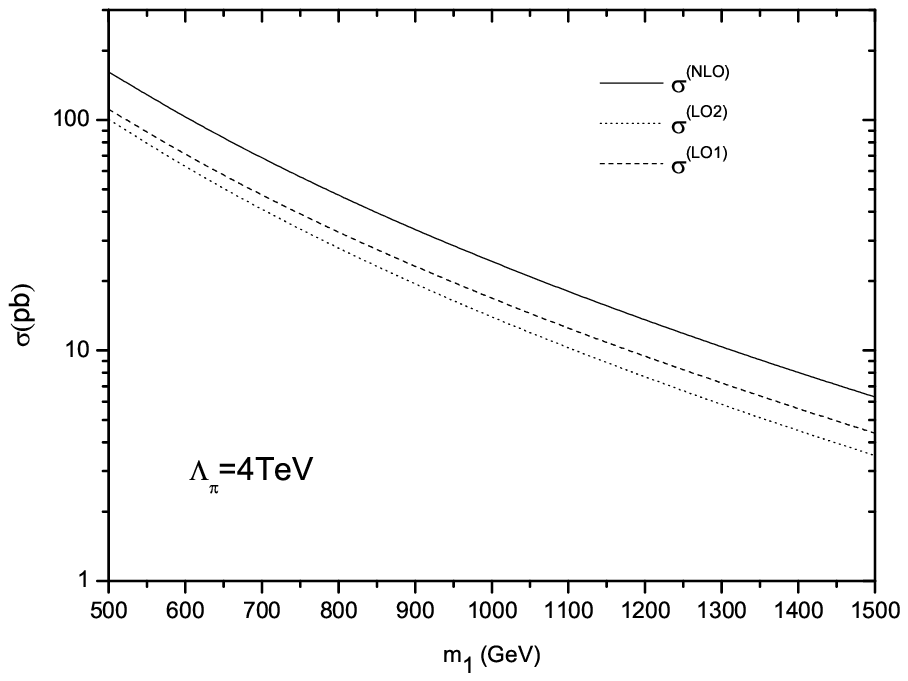, width=400pt}}
\caption[]{Dependence of the total cross sections for the first KK
graviton excitation mode direct production at the LHC on
$m_1$.}\label{9f7} \end{figure}

In Fig.~\ref{9f7}, we show the dependence of the total cross
sections of $pp\rightarrow h^{(1)}_{\mu\nu}$ at the LHC as functions
of $m_1$, assuming $\Lambda_\pi=4$\,TeV. The NLO and LO total cross
sections decrease when $m_1$ increases. The LO total cross sections
are in general over several pb, and reach 100 pb when
$m_1=500$\,GeV. Moreover, the figure also shows that the NLO QCD
corrections enhance significantly the LO total cross sections, which
are in general several tens percent.

\begin{figure}[ht!] \vspace{1.0cm} \centerline{\epsfig{file=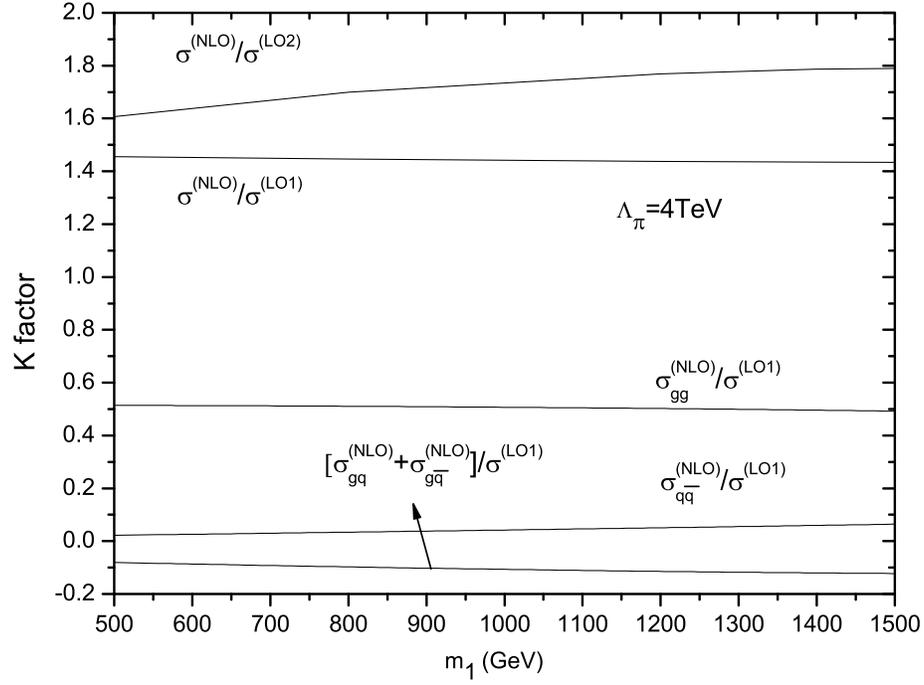,
width=400pt}} \caption[]{Dependence of the $K$-factor on $m_1$,
based on the results in Fig.\ref{9f7}.}\label{9f8} \end{figure}

Fig.~\ref{9f8} shows the dependence of the K factors on $m_1$, based
on the results in Fig.~\ref{9f7}. When $m_1$ varies from 500\,GeV to
1.5\,TeV, the $K_1$ factor ranges from 1.46 to 1.44, and the $K_2$
factor ranges from 1.61 to 1.71. In addition, we give the different
parts of the $K_1$ factor, which show that the contributions from
$\sigma_{gg}^{(NLO)}$ ranges from 0.51 to 0.44, the contributions
from $\sigma_{q\bar{q}}^{(NLO)}$ range from 0.03 to 0.14, and the
contributions from $\sigma_{gq}^{(NLO)}+\sigma_{g\bar{q}}^{(NLO)}$
range from -0.08 to -0.14, respectively.

\begin{figure}[ht!] \vspace{1.0cm}
\centerline{\epsfig{file=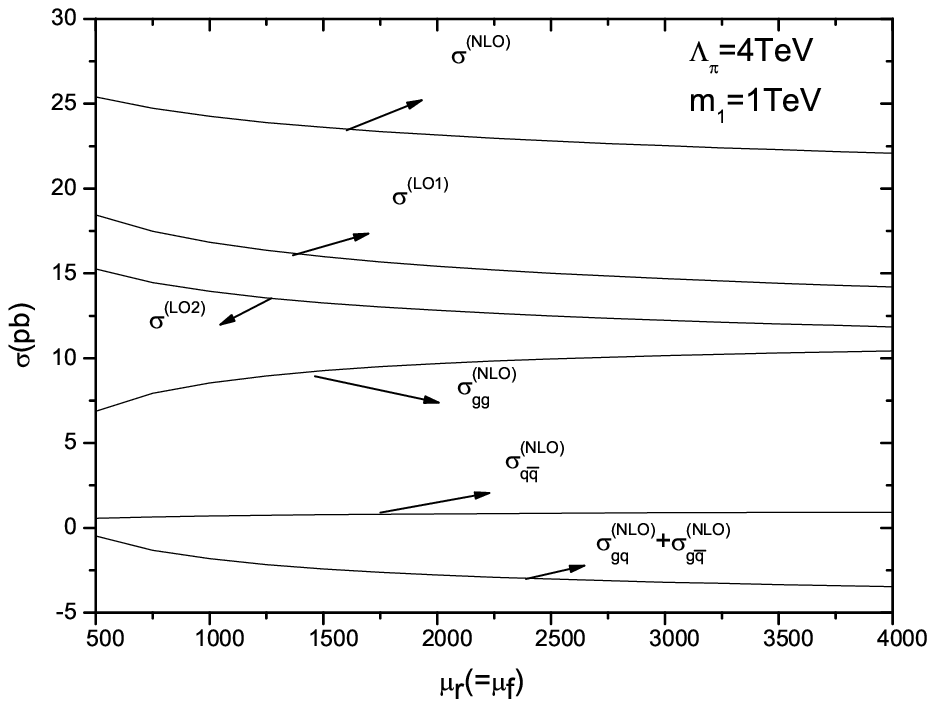, width=400pt}}
\caption[]{Dependence of the total cross sections for the first KK
graviton excitation mode direct production at the LHC on
$\mu_r=\mu_f$.}\label{9f9} \end{figure}

Fig.~\ref{9f9} shows the dependence of the total cross sections of
$pp\rightarrow h^{(1)}_{\mu\nu}$ at the LHC on the renormalization
scale ($\mu_r$) and the factorization scale ($\mu_f$), assuming
$\mu_r=\mu_f$, $\Lambda_\pi=4$\,TeV and $m_1=1$\,TeV. The scale
dependence of the NLO total cross section is much smaller than
that of the LO cross section. For example, the LO cross sections
$\sigma^{(LO1)}$ ($\sigma^{(LO2)}$) vary by $\sim\pm 17.8\%$
($\sim\pm 17.3\%$), when $\mu_r=\mu_f$ ranges between $500$\,GeV
and $4$\,TeV, while the NLO ones vary by $\sim \pm 9.3\%$.
Moreover, we also give the scale dependence of the different parts
in $\sigma^{(NLO)}$, for example, when $\mu_r(=\mu_f)$ ranges
between $500$\,GeV and $4$\,TeV, $\sigma_{gg}^{(NLO)}$ varies from
6.9pb to 10.4pb, $\sigma_{q\bar{q}}^{(NLO)}$ varies from 0.56pb to
0.93pb, and $\sigma_{gq}^{(NLO)}+\sigma_{g\bar{q}}^{(NLO)}$ varies
from -0.5pb to -3.5pb, respectively.

\begin{figure}[ht!] \vspace{1.0cm}
\centerline{\epsfig{file=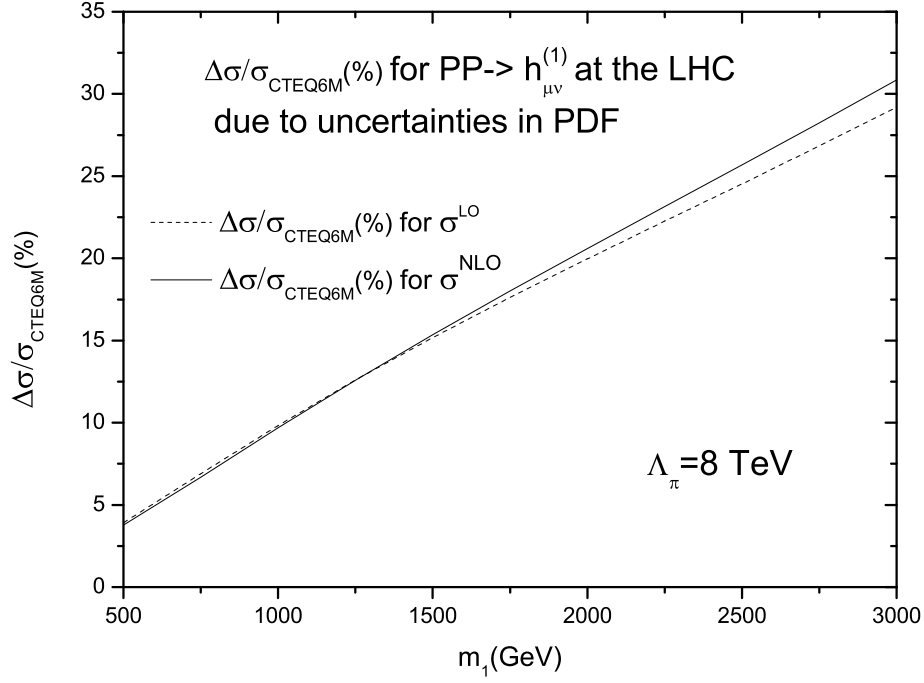, width=400pt}} \caption[]{The
PDF dependence of the total cross sections for the first KK
graviton excitation mode direct production at the LHC, as
functions of $m_1$.}\label{9f10}
\end{figure}

To estimate the uncertainties in the total cross sections due to the
uncertainty of PDFs, we take the 41 sets of CTEQ6.1 PDFs to
calculate the LO and NLO rates~\cite{61cteq}. Fig.~\ref{9f10} shows
the PDF uncertainties (defined here as the Eq.~(3) in
Ref.~\cite{PDFUU}) in the LO and NLO total cross sections for
$pp\rightarrow h^{(1)}_{\mu\nu}$ production at the LHC, as functions
of $m_1$, assuming $\Lambda_\pi=8$\,TeV. It turns out that the PDF
uncertainties in the LO and NLO total cross sections increases as
$m_1$ increases. Moreover, when $m_1$ is small ($<1.5$\,TeV), the
PDF uncertainties in the LO and NLO total cross sections are about
the same, while when $m_1$ becomes large ($>1.5$\,TeV), the NLO rate
has a larger uncertainty than the LO rate due to the PDF
uncertainties, especially at large $m_1$.

\begin{figure}[ht!] \vspace{1.0cm} \centerline{\epsfig{file=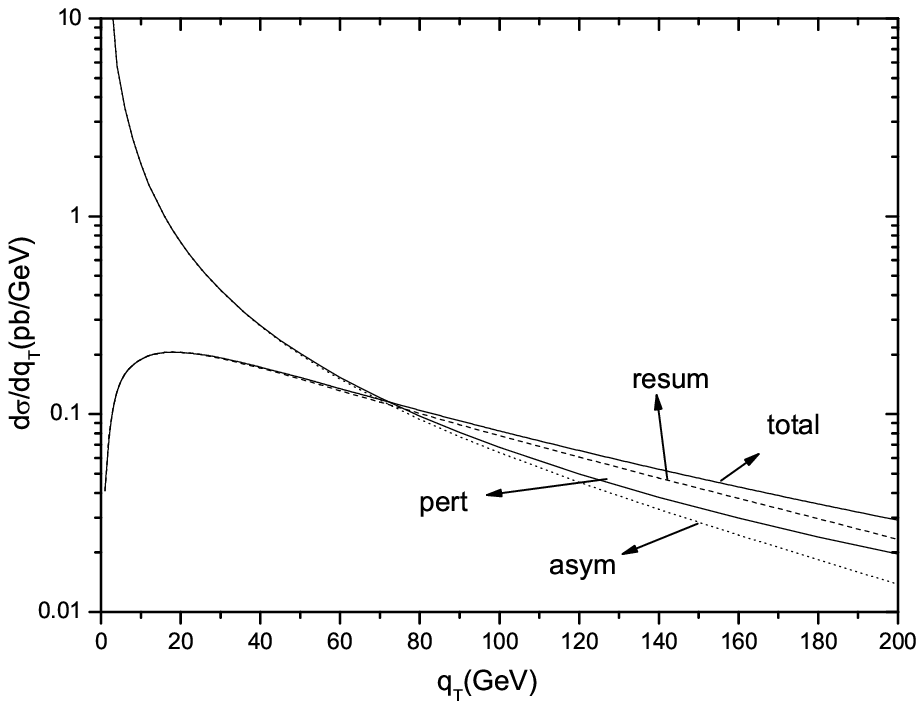,
width=300pt}} \caption[]{The transverse momentum distribution of
the first KK graviton excitation mode from $pp\rightarrow
h^{(1)}_{\mu\nu}$ process at the LHC, assuming
$\Lambda_\pi=4$\,TeV and $m_1=1$\,TeV.}\label{9f11}
\end{figure}

\begin{figure}[ht!] \vspace{1.0cm} \centerline{\epsfig{file=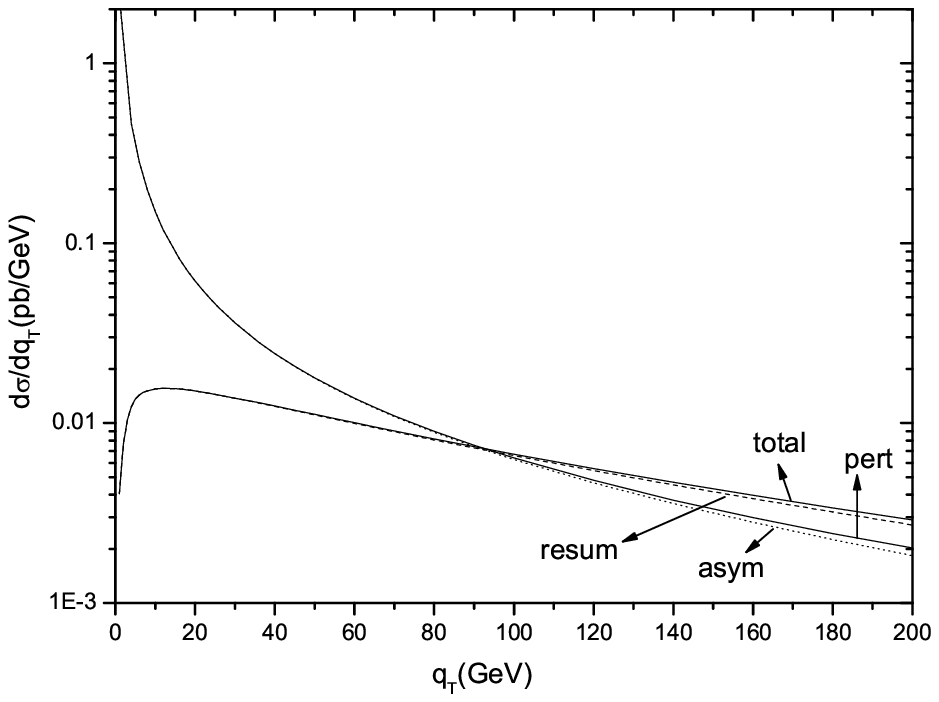,
width=300pt}} \caption[]{The transverse momentum distribution of
the first KK graviton excitation mode from $pp\rightarrow
h^{(1)}_{\mu\nu}$ process at the LHC, assuming
$\Lambda_\pi=4$\,TeV and $m_1=2$\,TeV.}\label{tot2000}
\end{figure}

Figs.~\ref{9f11} and \ref{tot2000} gives the transverse momentum
distribution of $h^{(1)}_{\mu\nu}$ from $pp\rightarrow
h^{(1)}_{\mu\nu}$ process at the LHC, assuming
$\Lambda_\pi=4$\,TeV, for $m_1=1$\,TeV and $2$\,TeV, respectively.
The peaks of the distribution appear at about $18$\,GeV and
$13$\,GeV. The differential cross sections decrease sharply with
the increase of $q_T$, which indicates that most events will
happen in the relatively low $q_T$ region, where the resummation
effects are essential. Moreover, we also plot the various parts of
the differential cross sections in Eq.(\ref{eq:total00}). The
perturbative and the asymptotic cross sections agree very well at
small transverse momentum. On the other hand, the resummed and the
asymptotic part are not canceled completely at high $q_T$ due to
the higher order effects included in the resummed one, so that the
total one and the perturbative one will differ at large $q_T$.
This can be considered as the theoretical uncertainties. In
principle, one can return to the perturbative result for
$q_T>q_T^{\text{cut}}$, where $q_T^{\text{cut}}$ is arbitrarily
chosen in the intermediate $q_T$ region.  However, in order to
make the transition smooth, one must introduce some kinds of
matching procedure which could also lead to uncertainties. In our
work, as shown by Eq.~(\ref{asyys}), we subtract from the Y term
the expansion of the resummed part of the same perturbative order,
and this matching procedure between small $q_T$ and large $q_T$
region prevents double-counting of perturbative results and also
guarantees a uniform theoretical accuracy over the entire $q_T$
region\cite{Klasen}.

\begin{figure}[ht!] \vspace{1.0cm}
\centerline{\epsfig{file=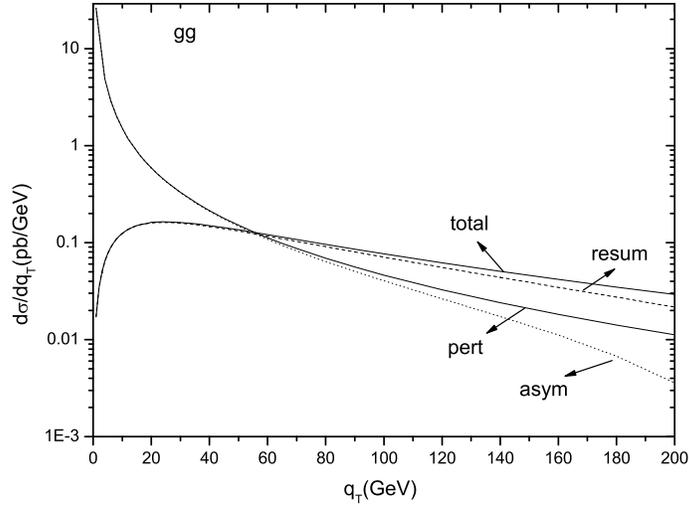, width=300pt}} \caption[]{The
$gg$ part of the transverse momentum distribution of the first KK
graviton excitation mode, assuming $\Lambda_\pi=4$\,TeV and
$m_1=1$\,TeV. }\label{9f12}
\end{figure}

\begin{figure}[ht!] \vspace{1.0cm}
\centerline{\epsfig{file=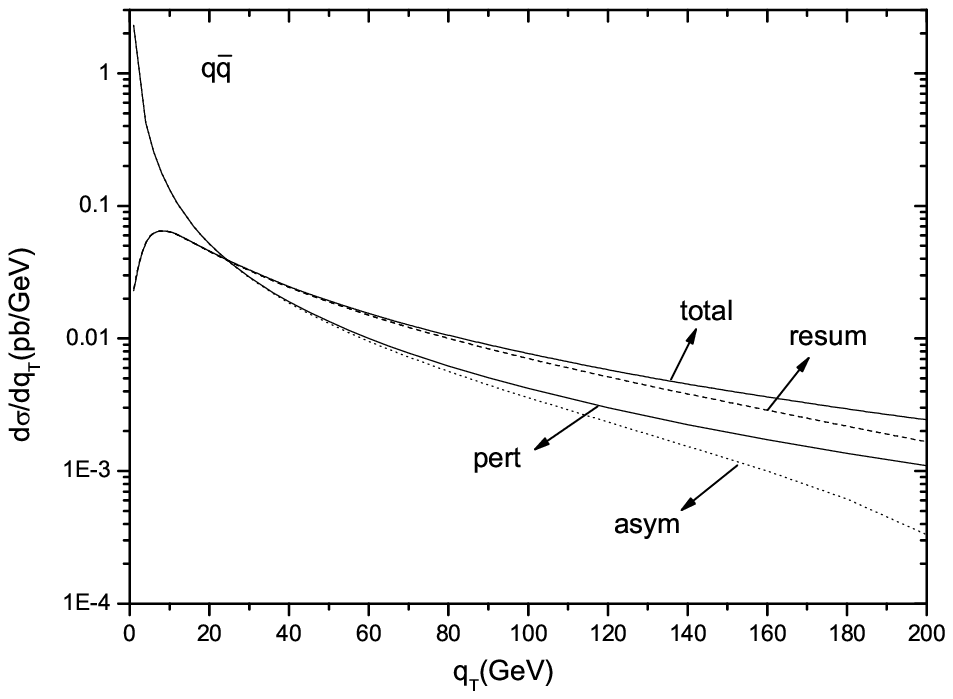, width=300pt}} \caption[]{The
$q\bar{q}$ part of the transverse momentum distribution of the
first KK graviton excitation mode, assuming $\Lambda_\pi=4$\,TeV
and $m_1=1$\,TeV. }\label{9f13}
\end{figure}

\begin{figure}[ht!] \vspace{1.0cm}
\centerline{\epsfig{file=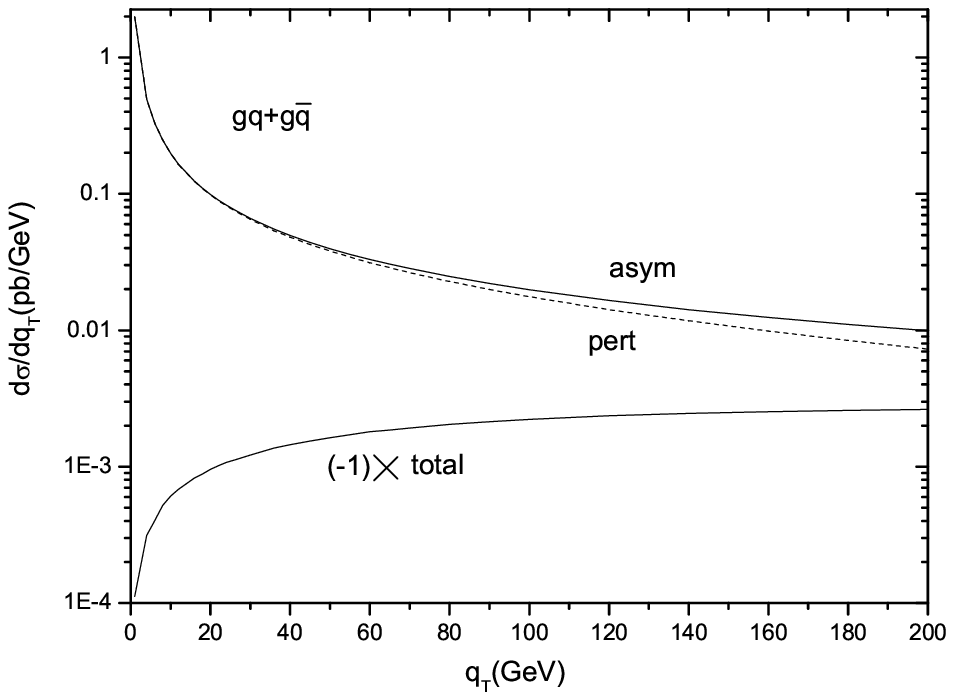, width=300pt}} \caption[]{The
$gq$ and $g\bar{q}$ part of the transverse momentum distribution
of the first KK graviton excitation mode, assuming
$\Lambda_\pi=4$\,TeV and $m_1=1$\,TeV. }\label{9f14}
\end{figure}

From Eqs.~(\ref{eq:total11}), (\ref{eq:total12}) and
(\ref{asyys}), we know that the transverse momentum distribution
can be divided into three parts, i.e. the contributions from $gg$,
$q\bar{q}$ and $gq+g\bar{q}$ channels:
\begin{eqnarray}
  \frac{d\sigma}{dq_T^2dy}({\rm total})=  \frac{d\sigma_{gg}}{dq_T^2dy}({\rm total})+
  \frac{d\sigma_{q\bar{q}}}{dq_T^2dy}({\rm total})+\frac{d\sigma_{gq+g\bar{q}}}{dq_T^2dy}({\rm total})
\end{eqnarray}
where
\begin{eqnarray} &&
\frac{d\sigma_{gg}}{dq_T^2dy}({\rm
total})\equiv\frac{d\sigma_{gg}}{dq_T^2dy}
  ({\rm resum})+\frac{d\sigma_{gg}}{dq_T^2dy} ({\rm pert}) -
  \frac{d\sigma_{gg}}{dq_T^2dy} ({\rm asym}) \\
&& \frac{d\sigma_{q\bar{q}}}{dq_T^2dy}({\rm
total})\equiv\frac{d\sigma_{q\bar{q}}}{dq_T^2dy}
  ({\rm resum})+\frac{d\sigma_{q\bar{q}}}{dq_T^2dy} ({\rm pert}) -
  \frac{d\sigma_{q\bar{q}}}{dq_T^2dy} ({\rm asym})\\
&&\frac{d\sigma_{gq+g\bar{q}}}{dq_T^2dy}({\rm total})\equiv
\frac{d\sigma_{g\bar{q}}}{dq_T^2dy} ({\rm pert})+\frac{d\sigma_{gq}}{dq_T^2dy} ({\rm pert})\nonumber\\
&&\hspace{3.1cm} -
  \frac{d\sigma_{g\bar{q}}}{dq_T^2dy} ({\rm asym})- \frac{d\sigma_{gq}}{dq_T^2dy} ({\rm asym})
\end{eqnarray}
In Figs.\ref{9f12}, \ref{9f13} and \ref{9f14}, we thus plot these
three contributions to the the transverse momentum distribution of
$h^{(1)}_{\mu\nu}$ based on the results in Figs.\ref{9f11},
respectively. We can see that for all these three parts, the
perturbative and the asymptotic cross sections agree very well at
small transverse momentum.

\section{Conclusions}
In summary, we have calculated the next-to-leading order total cross
section and transverse momentum distribution of single massive
graviton production at the LHC in the RS model, including all-order
soft gluon resummation effects. Our results show that the LO total
cross sections are in general over several pb in most of the
parameter space, and can reach 100 pb when $m_1=500$\,GeV. The NLO
corrections enhance significantly the total cross sections, which is
in general several tens percent, and reduce efficiently the
dependence of the total cross sections on the
renormalization/factorization scale. We have also examined the
uncertainty in total cross sections due to the PDF uncertainties,
and found that the uncertainty in NLO cross sections is slightly
larger than that in LO ones, especially at large $m_1$. For the
transverse momentum distribution, within the CSS resummation
formalism, we resum the logarithmically-enhanced terms at small
$q_T$ to all orders up to NLO logarithmic accuracy. Combined with
the fixed order calculations, we give consistent predictions for
both small $q_T$ and large $q_T$. Our results can be useful to the
simulation of the events in the future collider experiments.

\begin{acknowledgments}
We thank Yang Gao for useful discussion. This work was supported
in part by the National Natural Science Foundation of China,
  under grants No.~10421503 and No.~10575001, the Key Grant Project of Chinese Ministry
  of Education under grant No.~305001.
\end{acknowledgments}

\newpage


\begin{thebibliography}{99}
\bibitem{ADD} N. Arkani-Hamed et al.,
Phys. Lett. {\bf B429}, 263 (1998); N. Arkani-Hamed et al., Phys.
Rev. {\bf D59}, 086004 (1999); N. Arkani-Hamed et al., Phys. Lett.
{\bf B436}, 257 (1998).
\bibitem{RS} L. Randall, R.Sundrum, Phys. Rev. Lett. {\bf 83}, 3370 (1999);
L. Randall, R.Sundrum, Phys. Rev. Lett. {\bf 83}, 4690 (1999).
\bibitem{lyk} J. Lykken, Phys. ReV. {\bf D54}, R3693 (1996).
\bibitem{witt} E. Witten, Nucl. Phys. {\bf B471}, 135 (1996).
\bibitem{hora} P. Horava and E. Witten, Nucl. Phys. {\bf B460}, 506 (1996);
 Nucl. Phys. {\bf B475} (1996) 94.
\bibitem{anto} I. Antoniadis, Phys. Lett. {\bf B246}, 377 (1990).
\bibitem{csaki}C. Csaki, TASI Lectures on Extra Dimensions and
Branes, hep-ph/0404096.
\bibitem{grav} E. G. Adelberger [Eot-Wash Group Collaboration] hep-ex/0202008.
\bibitem{Hewett1} J.L. Hewett, Phys. Rev. Lett. {\bf 82} 4765 (1999).
\bibitem{9909255}H. Davoudiasl, et al., Phys.Rev.Lett. {\bf 84}, 2080 (2000).
\bibitem{0205106}JoAnne Hewett, Maria Spiropulu, Ann.Rev.Nucl.Part.Sci. {\bf52}, 397 (2002).
\bibitem{0506158} Prakash Mathews et al., JHEP {\bf0510}, 031 (2005);
Prakash Mathews et al., Nucl.Phys. {\bf B713}, 333 (2005); Prakash
Mathews, V. Ravindran, hep-ph/0507250;  M. C. Kumar et al.,
hep-ph/0604135.
\bibitem{0211205} B. C. Allanach, et al., JHEP {\bf
0212}, 039 (2002).
\bibitem{Hant}
T. Han et al., Phys. Rev. {\bf D59}, 105006 (1999).
\bibitem{beenakker1} W. Beenakker et al., Phys. Rev. {\bf D40}, 54 (1989).
\bibitem{DREG} G.~'t Hooft, M.~J.~G.~Veltman, Nucl. Phys. {\bf B44}, 189 (1972).
\bibitem{altarelli} G.~Altarelli et al.,
Nucl. Phys. {\bf B157}, 461 (1979); J.~C.~Collins et al., in:
Perturbative Quantum Chromodynamics, ed. A.~H.~Mueller (World
Scientific, 1989).
\bibitem{cutoff} B.~W.~Harris, J.~F.~Owens, Phys. Rev. {\bf D65}, 094032 (2002).
\bibitem{alpha}
G.~Altarelli, G.~Parisi, Nucl. Phys. {\bf B126}, 298 (1977).
\bibitem{Nucl.Phys.B250.199}
  J.~C.~Collins, D.~E.~Soper and G.~Sterman,
  Nucl. Phys. {\bf B250}, 199 (1985).
  \bibitem{hhgres}R.P. Kauffman, Phys. Rev. {\bf D44}, 1415 (1991).
\bibitem{dygres}P.B. Arnold, R.P. Kauffman, Nucl. Phys.
{\bf B349}, 381 (1991); C. Bal{\'{a}}zs et al., Phys. Lett. {\bf
B355}, 548 (1995).
\bibitem{Phys.Rev.D67.073016}
  F.~Landry et al.,
  Phys. Rev. {\bf D67}, 073016 (2003).
  \bibitem{CTEQ} J.~Pumplin et al., JHEP {\bf 0207}, 012 (2002).
  \bibitem{61cteq} D.~Stump et al., JHEP {\bf 0310}, 046 (2003).
\bibitem{PDFUU} J. Pumplin et al., JHEP 0207, 012(2002).
\bibitem{Klasen} G. Bozzi, B. Fuks and M.
Klasen, Phys. Rev.{\bf D74}, 015001 (2006).
\end{thebibliography}
\end{document}